\documentclass[aps,pre,reprint,groupedaddress]{revtex4-1}
\usepackage{graphicx}
\usepackage{graphics}

\usepackage{epsf,wrapfig,epsfig}

\begin{document}

%\preprint{}

\title{Demonstration of Experimental Three Dimensional Finite Time Lyapunov Exponents with Inertial Particles}


%\author{}
%\email[]{sraben@vt.edu}
%%\homepage[]{Your web page}
%%\thanks{}
%%\altaffiliation{}
%\affiliation{Virginia Tech}
%\author{Shane D. Ross}
%\email[]{sdross@vt.edu}
%%\homepage[]{Your web page}
%%\thanks{}
%%\altaffiliation{}
%\affiliation{Virginia Tech}
%\author{Pavlos P. Vlachos}
%\email[]{pavlos.vlachos@gmail.com}
%%\homepage[]{Your web page}
%%\thanks{}
%%\altaffiliation{}
%\affiliation{Purdue University}

\author{Samuel G. Raben\footnote{Virginia Tech}, \
  Shane D. Ross\footnotemark[1],\ 
  Pavlos P. Vlachos\footnote{Purdue University}
  }


\date{\today}

\begin{abstract}
This work provides an experimental method for simultaneously measuring finite time Lyapunov exponent fields for multiple particle groups, including non-flow tracers, in three-dimensional multiphase flows.  From sequences of particle images, e.g., from experimental fluid imaging techniques, we can directly compute the flow map and coherent structures, with out performing the computationally costly numerical integration.  This is particularly useful to find three-dimensional Lagrangian coherent structures for inertial particles, that do not follow the bulk fluid velocity, as we demonstrate for a grid turbulence experiment.  The technique described may provide a new means for exploring the physics of experimental multi-phase flows.
\end{abstract}

%\keywords{}

\maketitle

%% body of paper here - Use proper section commands
%% References should be done using the \cite, \ref, and \label commands
%\section{}
%% Put \label in argument of \section for cross-referencing
%%\section{\label{}}
%\subsection{}
%\subsubsection{}

Finite Time Lyapunov Exponents (FTLE) are a powerful and increasingly popular tool for describing mixing and transport in both turbulent and laminar flow fields \cite{Haller2001, Brunton2010}.  FTLEs provide a measure of the exponential rate of divergence or convergence of Lagrangian particle trajectories.  They can be used both experimentally and numerically to describe a flow field, which may have a high degree of spatiotemporal complexity  \cite{Haller2001, Shadden2007, Shadden2006}.  While FTLEs are primarily used to describe single-phase flow behavior \cite{Haller2001, Haller2005, Shadden2006} some works have attempted to account for inertial particles by modeling the particles' motion through simulations \cite{Haller2008, Tallapragada2008, Peng2009}.  This procedure can provide insight, but does not provide direct information about the true observable inertial particle trajectories.  Often, the equations for inertial particle motion make simplifying assumptions (e.g., the Maxey-Riley equations \cite{Maxey1983}) that can lead to significant differences between the modeled and true motion.  This brief communication describes a method to directly determine FTLEs from experimental data for inertial particles through the use of particle tracking velocimetry (PTV) without any {\it a-priori} assumptions about particle motion.  

FTLEs are computed via the Cauchy-Green deformation tensor $C_{jk}$,
\begin{equation}
C=\left(\nabla \Phi^{t_0+T}_{t_0}\right)^* \cdot \nabla \Phi^{t_0+T}_{t_0}
\end{equation}
where * denotes the matrix transpose, and $\Phi^{t_0 + T}_{t_0}$ is the flow map (diffeomorphism) of particle locations from time $t_0$ to $t_0+T$, where $T$ is the time over which the FTLEs will be computed.  From the maximum eigenvalue of C, the FTLE field defined in the measurement volume is,
\begin{equation}
\sigma^{t_0+T}_{t_0} = \frac{1}{T} \ln \left( \sqrt{\lambda_{max} \left(C\right)}\right)
\end{equation}
typically when computing FTLEs from experimental data to use a numerical integration routine to numerically advect artificial tracer particles to determine the flow map from the estimated velocity field \cite{Shadden2006, Shadden2007, Lekien2010}.  While this can be effective for single-phase flow it neglects the fact that inertial particles, bubbles, or active particles may fail to follow the bulk fluid motion or the fact that tracking individual particles can provide a direct measure of a short duration flow map.  Lagrangian tracking can provide a measure of the flow map over longer times but is more prone to experimental errors \cite{Raben2013}.  While numerical routines can be modified to estimate the inertial particle behavior via modeling as mentioned above, this procedure does not directly measure inertial particle trajectories.  However, using time resolved PTV to obtain snap shots of the particle motion allows direct measurement of the particle flow map while also allowing for parameterization of the particle flow map based on unique identifying characteristics, such as size, shape or color, providing, e.g., a one-parameter family of particle flow maps with particle size as the parameter.  The concept of merging small flow map snap shots to estimate a complete flow map was put forth by Brunton and Rowley \cite{Brunton2010} for results of fluid computations and later adapted for experimental data as PTV interpolation by Raben et al. \cite{Raben2013}.  Through this method it is possible to simultaneously determine FTLEs for multiple particle groups within the same measurement volume and compare them to the bulk flow field.  It has also been shown that this method can provide high accuracy flow map computation results even when the particle concentration drops below what is typically used for PIV/PTV \cite{Raben2013}.  This is an important aspect; when the particles are separated into groups, some groups will have smaller particle population densities requiring a method suitable to provide high resolution FTLE information with low resolution velocity information in order to properly determining the FTLE values.

To study the motion of inertial particles in an experimental environment, data were collected in a vertical water tunnel that was designed to generate homogeneous isotropic grid turbulence, as described in \cite{Raben2012}.  For this experiment a bar thickness of the grid, $b=0.3175$ cm was used with the gap between bars equal to the width of the bar.  Overlapping bars created a square lattice, which was located 8 cm upstream from the measurement location.  Two different types of particles where added to the flow: $85 \pm 20$ $\mu$m diameter silver coated hollow glass spheres that were tuned to be neutrally buoyant and were used to act as flow tracers; and solid glass particles with diameters ranging from approximately 150 - 200 $\mu$m that were added downstream (top of the tunnel) and had an approximate mass density of 2600 kg/m$^3$.  The vertical nature of the tunnel created opposing motion as gravity pulled the negatively buoyant particles down while the bulk flow was moving mostly upward.

Time resolved imaging techniques such as particle image velocimetry (PIV) have made it possible to study the Lagrangian motion of a flow field experimentally \cite{Mathur2007, Shadden2007}.  With the recent development of volumetric image techniques \cite{Elsinga:2006wo} it is now possible to investigate particle trajectories in a fully three-dimensional flow field.  Because these imaging techniques make no assumptions on particle motion (e.g., must be a tracer following the bulk flow) they can be effective in capturing non-flow tracer particle motion (e.g., inertial particles) as well as bulk flow motion.

Time resolved tomographic imaging was used to collect information on the complete particle field as well as fully resolve the three-dimensional fluid motion.  A New Wave Pegasus laser was used to illuminate all the particles in the measurement volume.  Three Photron FASTCAM APX-RS high-speed CMOS cameras were used to simultaneously image this light field, recording images at 250 Hz.  These images were reconstructed into a three-dimensional light intensity distribution using the Multiplicative Algebraic Reconstruction Technique (MART) employ in the DaVis 8.1 software \cite{Herman1976, Elsinga:2006wo}.

Once the images had been reconstructed, the particles' size and motion were determined.  Particles were first located in the volume using a simple thresholding method and then sized using an intensity weighted pixel count.  In an effort to track the particles, a multi-component particle tracking algorithm developed for single and multiphase flows \cite{Cardwell2011} was adapted to three-dimensional data.  The method worked by comparing unique particle identifiers, such as size, peak intensity, and proximity, to match particles in consecutive images.  This method has been shown to work well in turbulent flows even with non-flow tracers \cite{Cardwell2011}.

%-----START FIGURE-------
\begin{figure}[!]
\begin{tabular}{c}
\includegraphics[width=0.3\textwidth]{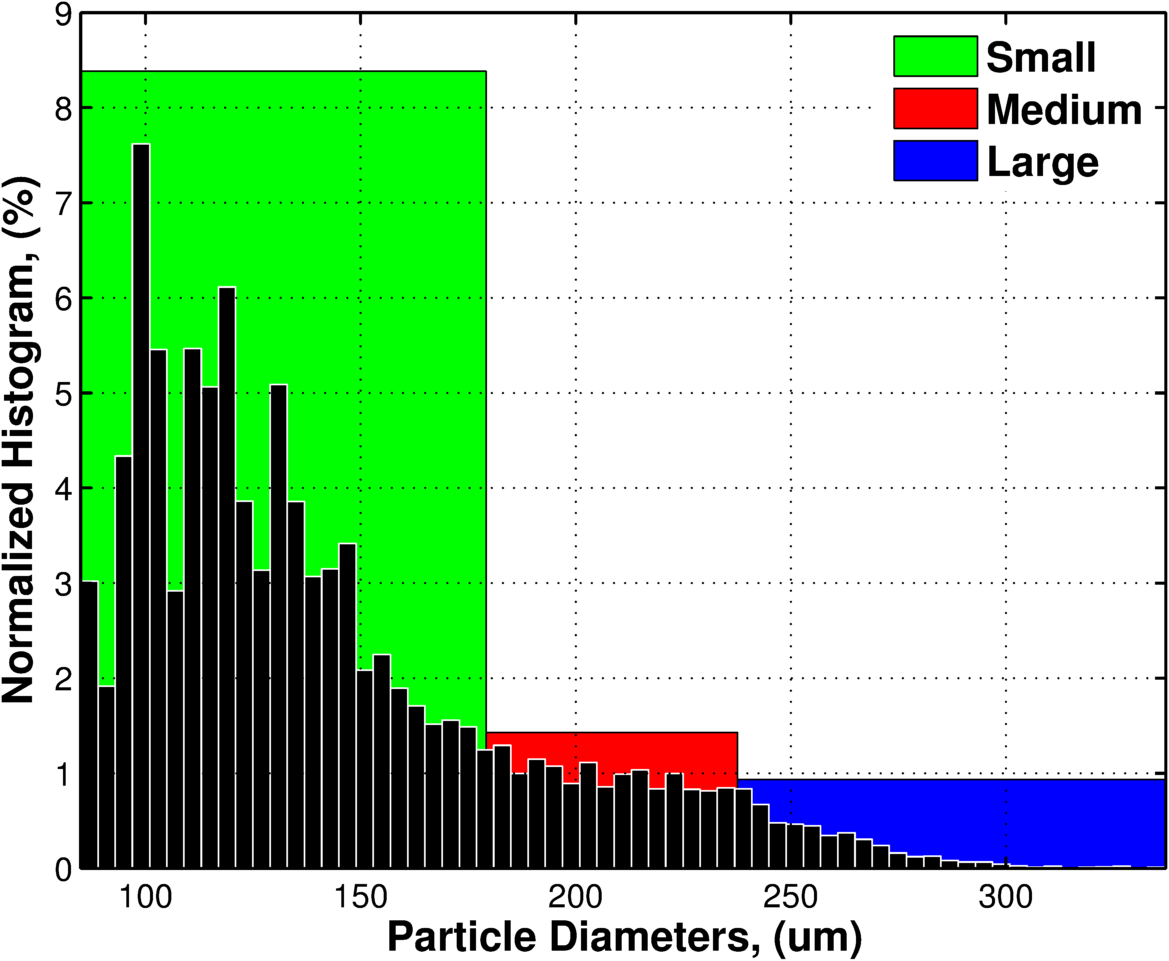} \\
{\footnotesize A} \\
\includegraphics[width=0.4\textwidth]{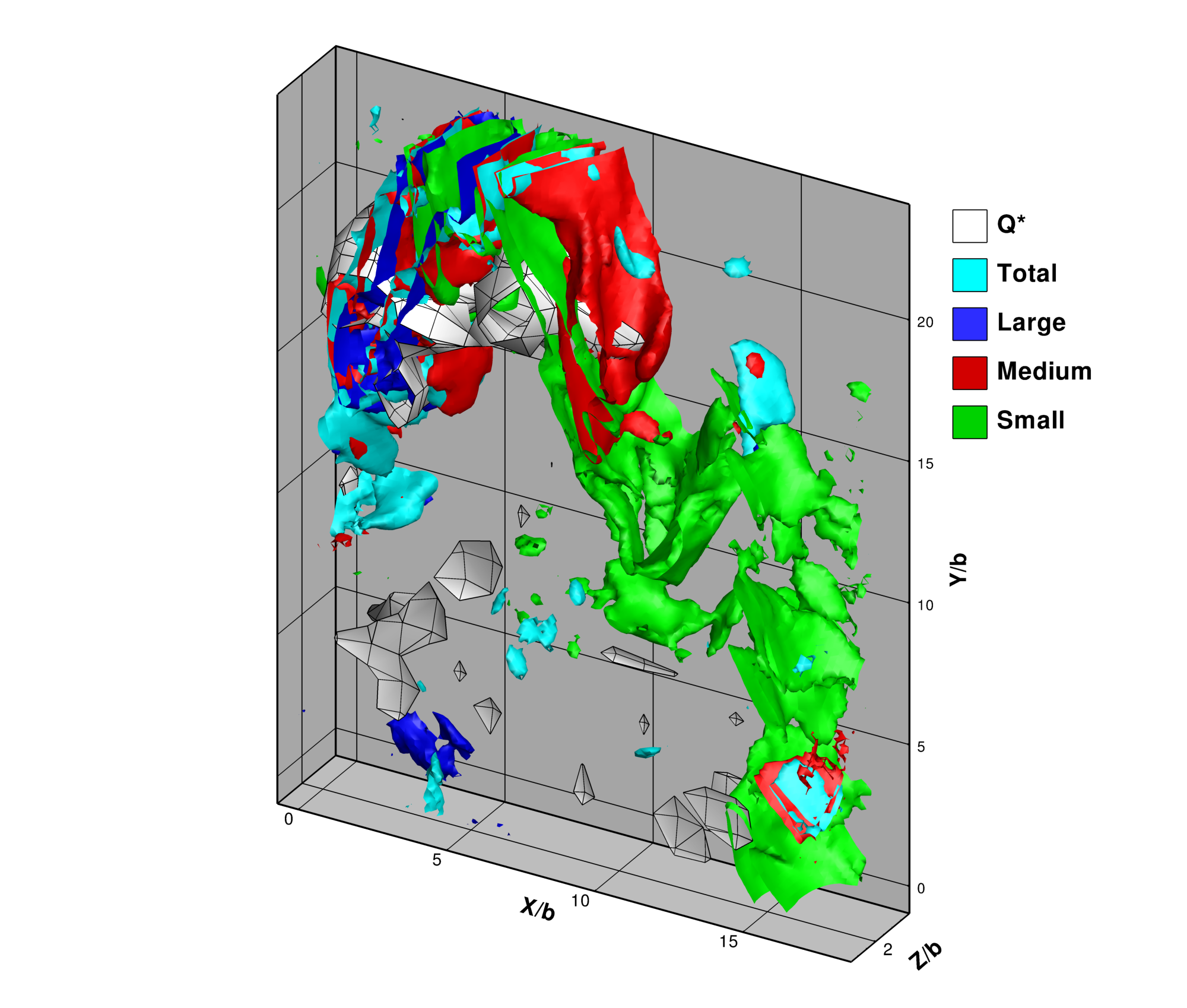}\\
{\footnotesize B} \\ 
\includegraphics[width=0.4\textwidth]{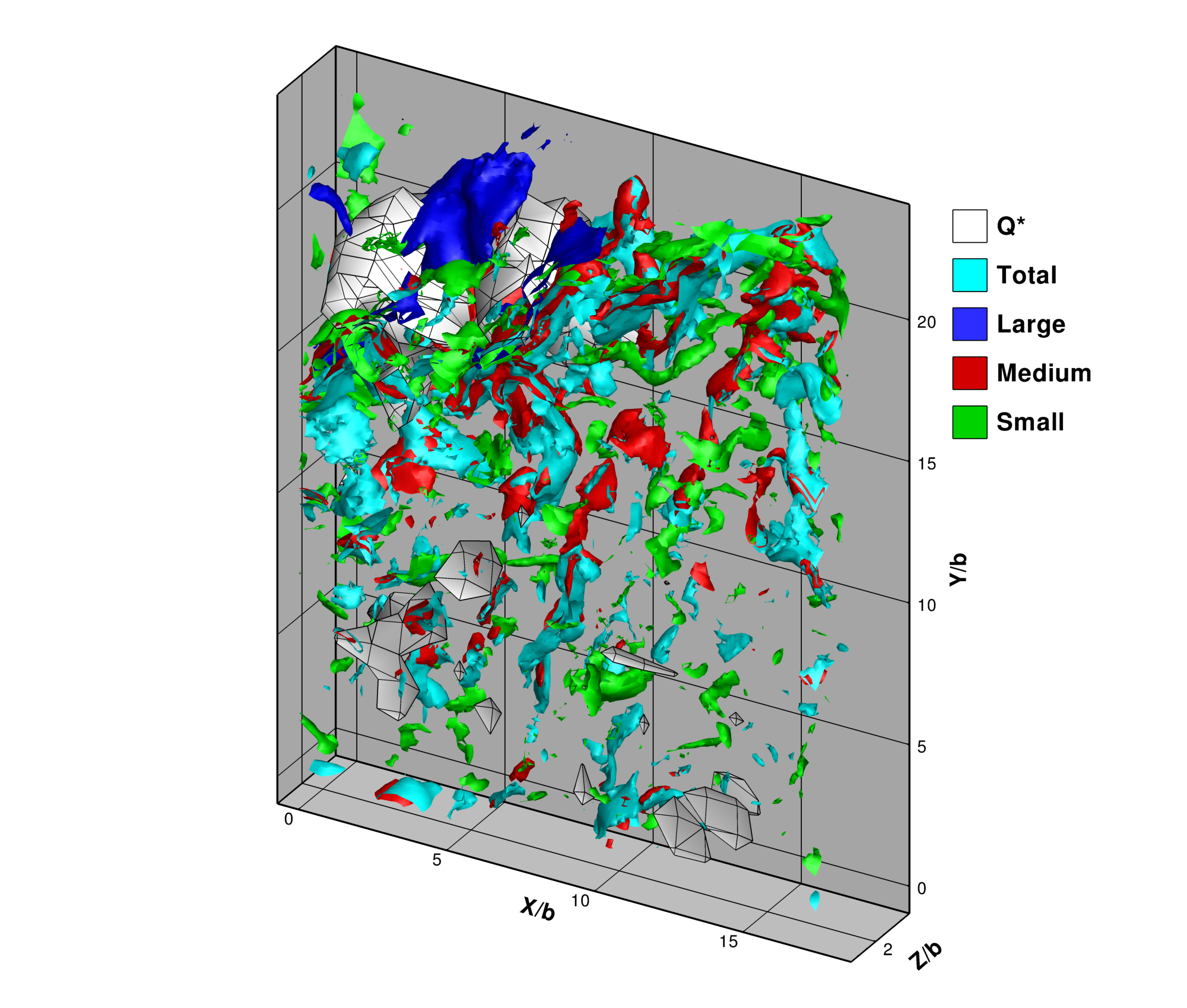}\\
{\footnotesize C} 
\end{tabular}
\caption{\label{fig:hist}(color online) (A) Normalized particle diameter distribution within the measurement volume. Iso-surfaces of the forward (B) and backward (C) FTLE fields based on the different components in the flow.
}
\end{figure} 
%-----END FIGURE-------

Figure \ref{fig:hist}A shows a histogram of the particle sizes present in the measurement volume.  Due to factors such as camera arrangement and the MART reconstruction algorithm \cite{Herman1976}, the particle size may be over-estimated.  As these factors should affect all particles equally, and the concern here is not exact particle size but rather relative size, this should not affect the results.  For this study, the particle size distribution was divided into only three groups.  The first group was composed of the smallest particles, most likely including the tracer particles, which should follow the bulk fluid motion.  The second group was composed of the medium particles, which contained a mixture of flow tracers and smaller glass particles.  The final group included the largest particles, which were primarily the large glass particles that will tend not to follow the bulk fluid motion.  When computing the FTLE field, the complete particle distribution was used as a control, as this total group provides an estimate of the FTLE field that would be found if no particle sizing procedure had been applied to the data and all the particles were (erroneously) treated as flow tracers.

The FTLE field was calculated for each particle group with an integration time of 1s which is equal to 250 frames.  For two-dimensional flows, FTLE fields are often characterized by the elevated ridges, or connected lines with high FTLE values, which are referred to as Lagrangian coherent structures (LCS) and reveal hyperbolic or shear-dominated structures.  In three-dimensional fields, the locus of elevated values are two-dimensional surfaces.  Figure \ref{fig:hist}B and C shows iso-surfaces of high FTLE values as proxies for true ridges for both the forward and backward FTLE fields.  Ridges in the forward FTLE field reveal repelling surfaces where particles are exponentially diverging away from one another while the backward FTLE shows attracting surfaces which may be related to clustering cores for inertial particles.  From Figure \ref{fig:hist}B it can be seen that there is a significant difference in the FTLE fields based on the particle size.  The iso-surface for the large particle group is dominated by a large structure in the upper left of the domain.  It could be seen from the raw data that during this time that there was an influx of larger particles that begin to spread throughout the volume, which would explain the elevated FTLE values in this region.  For the small particle group the iso-surface shows a structure that extends from the lower right of the domain up to the top.  This structure could indicate that the influx of large particles forced the flow tracers to be redirected around the large particle cluster causing a divergence in the small particle trajectories.

Figure \ref{fig:hist}C shows the backward FTLE, which will indicate locations of particle clustering.  Previous works that have investigated particle clustering have used the second invariant of the velocity gradient tensor, Q, sometimes referred to as the Okubo-Weiss parameter, as an indicator for where particles are likely to concentrate, \cite{Squires1991, Eaton1994, Guala2008, Haller2008} where Q is defined as,
\begin{equation}
Q=\frac{1}{2}\left( \omega^2 - s^2 \right)
\end{equation}
with $\omega$ and $s$ representing vorticity and strain rate, respectively.  For scaling purposes $Q$ is often normalized by the ensemble average of vorticity squared, $Q^*= Q/\left< \omega \right>$, as was done here.  This produced normalized values between -1.5 and 0.5 which is in agreement with the literature for turbulent flow \cite{Guala2008}.  When $Q^*$ is negative this indicates a region of high strain and low vorticity, which, when particles are added to the flow, has been shown to correlate with preferential particle concentration \cite{Squires1991, Eaton1994, Guala2008, Haller2008}.  To illustrate regions where particles should cluster a $Q^*$ iso-surface showing the location of three standard deviations away from the zero in the negative direction based on the mean field, is also shown in Figure \ref{fig:hist}.  It can be seen from Figure \ref{fig:hist}C that while there exist some smaller regions of high backward FTLE throughout the domain, the attracting LCS locations are predominantly located near the location of higher negative $Q^*$.  Since the flow is time-dependent, there is no reason to expect perfect agreement between the Eulerian $Q^*$ field and attracting LCSs.

%-----START FIGURE-------
\begin{figure*}[!]
\begin{tabular}{cc}
\includegraphics[width=0.31\textwidth]{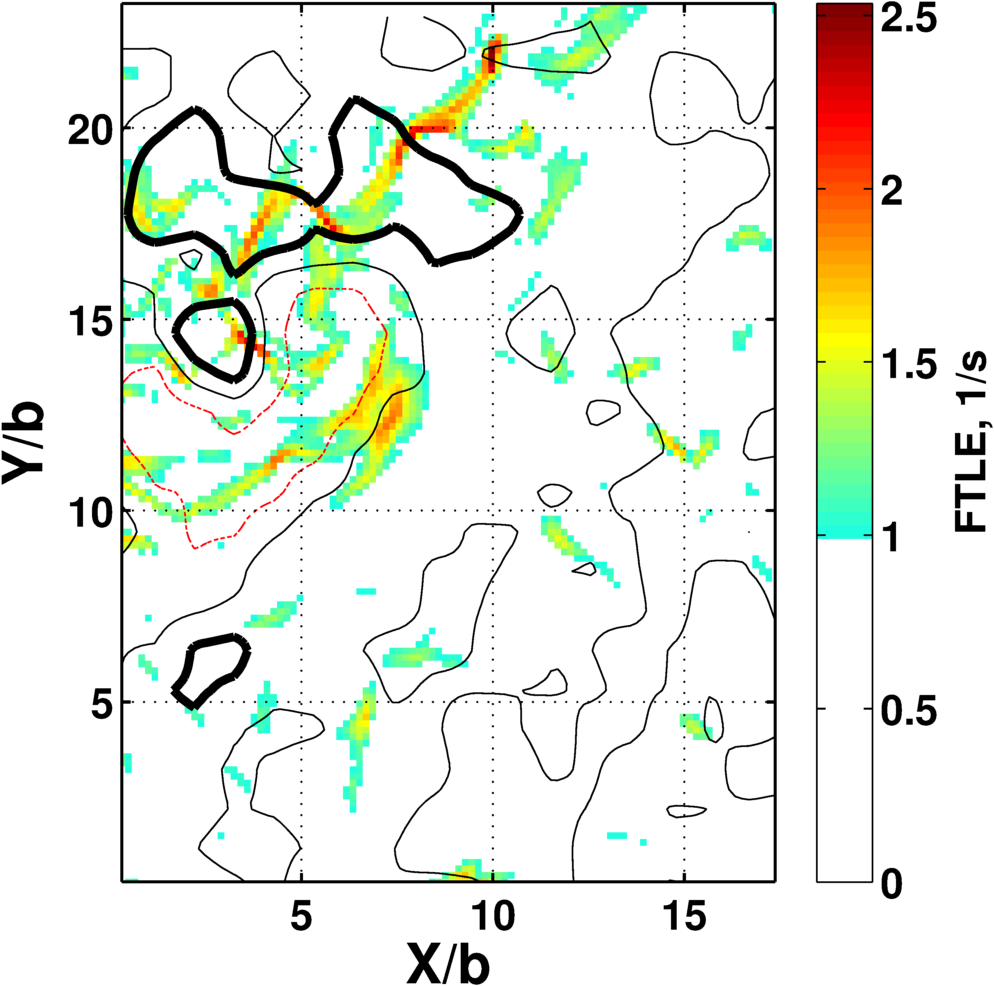}&
\includegraphics[width=0.31\textwidth]{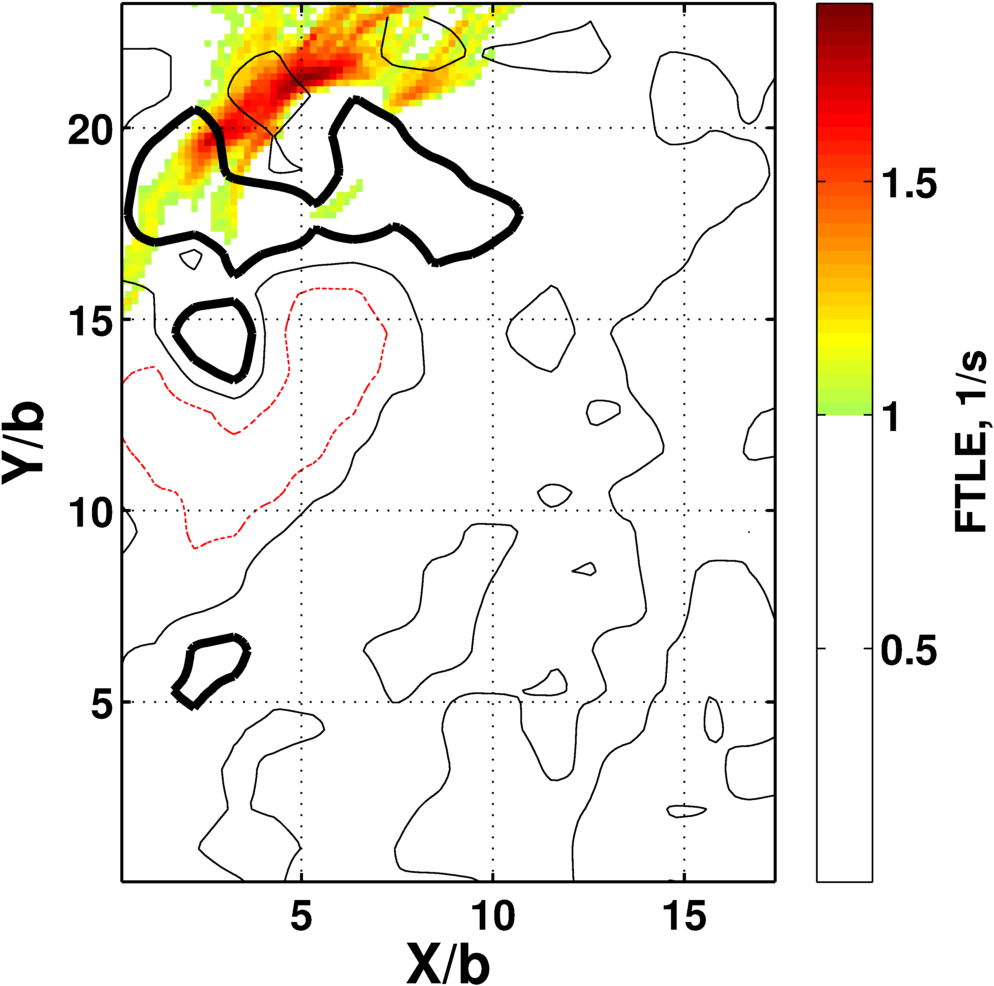}\\
{\footnotesize A} & 
{\footnotesize B} \\
{~} & {~} \\
\includegraphics[width=0.31\textwidth]{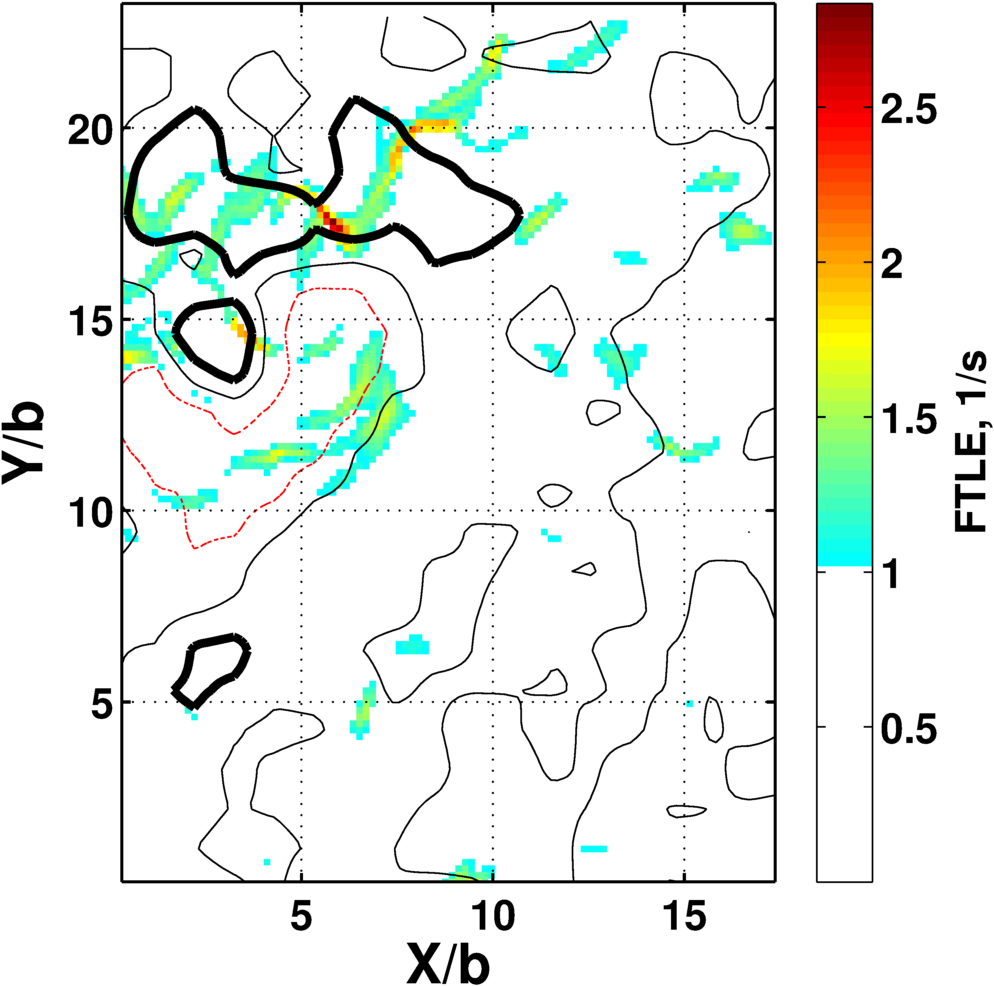}&
\includegraphics[width=0.31\textwidth]{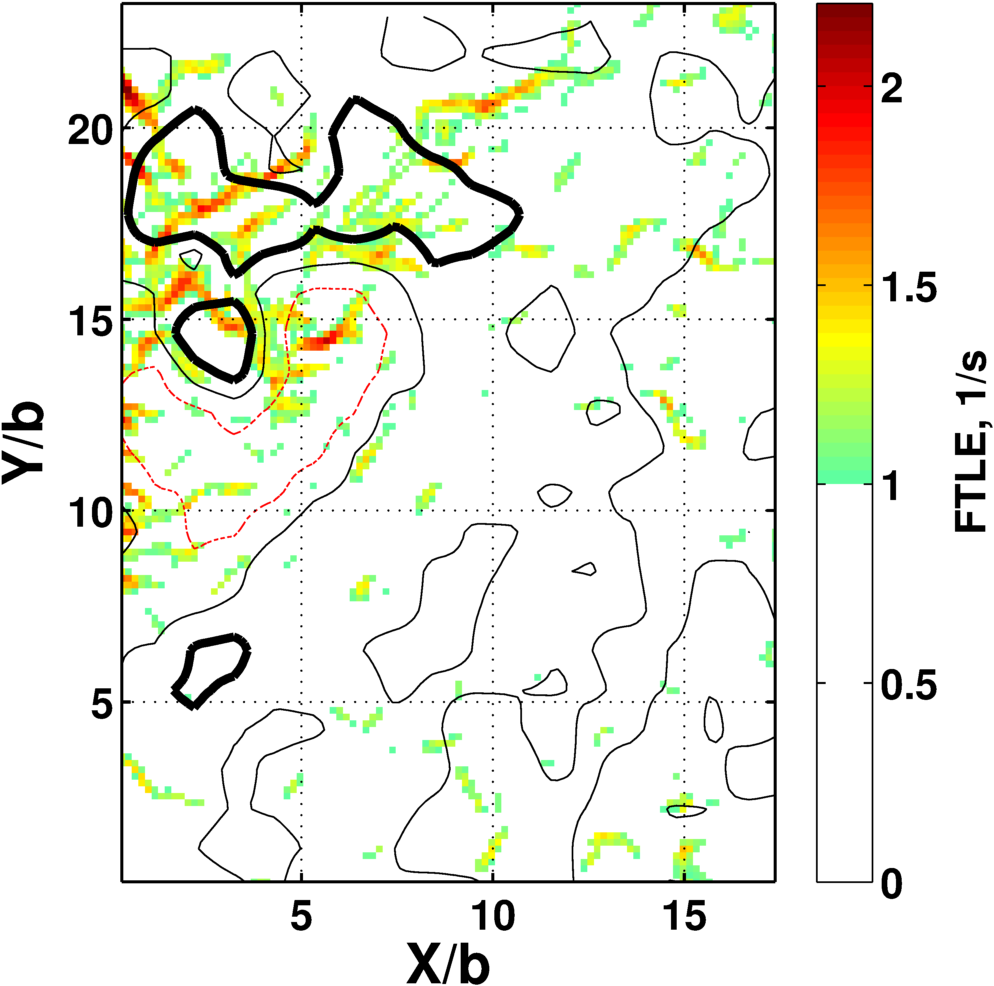}\\
{\footnotesize C} &
{\footnotesize D}
\end{tabular}
\caption{\label{fig:slice} (color online) Contours of Backward time FTLE values for the total location of particles A), the small particle sizes B) the medium particles sizes C) and finally the large particle sizes, D).  The thick line shows iso-contour for -3 std (surface shown in Figure \ref{fig:hist}B and C), while the thick line shows an iso-contour for -1.5 std and the dashed line is the zero iso-contour.}
\end{figure*}
%-----END FIGURE-------

To further investigate the locations of particle clustering, Figure \ref{fig:slice} shows backward FTLE values on the center Z plane for each of the 3 different particle groupings along with the total particle collection, with a thick black line representing the same iso-contour of $Q^*$ is included.  In addition an iso-contour -1.5 times the standard deviation and a zero contour are also included.  It can be seen from this figure that while there are some similarities in the locations of the elevated backward FTLE values between the different groups, there are also some important differences.  Figure \ref{fig:slice}A shows the FTLE field for the total particle group, which we note is not a superposition of the FTLE field for the size-based groups.  Elevated FTLE values are seen in close proximity to the highly negative $Q^*$ values as this will be a location where particles will cluster \cite{Guala2008}.  For the large particles, Figure \ref{fig:slice}D, elevated values are again seen near highly negative $Q^*$ but in a different location from that seen with the total particle group.  In this case the large particles appear attracted to a region just above the $Q^*$ iso-contour, on the opposite side from zero $Q^*$ iso-contour (the zero iso-contour would suggest particle repulsion).  The large particles also have a lower maximum FTLE value, which may indicate that their attraction to this region is not as strong as some of the other particles groups.

The medium particle group also has elevated FTLE values in close proximity to the $Q^*$ strongly negative iso-contour, as seen in Figure \ref{fig:slice}C.  As this group is most likely a collection of flow tracers and smaller inertial particles it is interesting to see that very high FTLE values appear to be located inside the $Q^*$ iso-contour mean that particle clustering associated with this group most closely coincides with the $Q^*$ grouping.  For the smallest particles, Figure \ref{fig:slice}B, it can again be seen that the elevated FTLE values are located near the $Q^*$ iso-contour.  This particle group appears to have more scatter than the other groups which is mostly due to the fact that as flow tracers these particles are more susceptible to the turbulent fluctuations in the volume and thus will have a more spatially distributed structure.  Again, because $Q^*$ is an Eulerian field and ours is a temporally varying flow there is no expectation of perfect agreement with the LCS but it does help to illustrate the behavior.

To summarize, this work has shown that three-dimensional FTLE fields can be calculated for inertial particles in experiments through the use a non-flow tracer flow map determination technique that uses particle tracking and sizing information to directly measure the size-parameterized families of flow maps.  The use of particle tracking for the direct calculation of the FTLEs is an important advancement as it is capable of uniquely determining the flow maps for different groups of particles, e.g., grouped by size in our experiment, but other parameterizations are possible.  Using this method it is possible to directly measure inertial particle FTLE fields and Lagrangian coherent structures without making assumptions about the underlying particle equations of motion.  This will have relevance for the experimental study of inertial particle motion in fluids and multi-phase flows.

SDR gratefully acknowledges partial support from NSF Grant 1150456.  

\bibliography{library.bib}

\end{document}